\documentclass[twocolumn,showpacs]{revtex4}

\usepackage{graphicx}
\usepackage{dcolumn}
\usepackage{bm}
\usepackage{amsmath}
\usepackage{amssymb}
\usepackage{graphics}

\begin{document}

\title{Experimental Tests of the New Paradigm for Laser Filamentation in Gases}

\author{Pavel Polynkin,$^{*}$ Miroslav Kolesik, Ewan~M.~Wright, and Jerome~V.~Moloney}
\affiliation{College of Optical Sciences, The University of Arizona, Tucson, Arizona 85721, USA}

\date{\today}

\begin{abstract}
Since their discovery in the mid-1990s, ultrafast laser filaments in gases have been described as products 
of a dynamic balance between Kerr self-focusing and defocusing by free electric charges that are 
generated via multi-photon ionization on the beam axis. 
This established paradigm has been recently challenged by a suggestion that 
the Kerr effect saturates and even changes sign at high intensity of light, and that this sign reversal, not
free-charge defocusing, is the dominant mechanism responsible for the extended propagation of laser filaments. 
We report qualitative tests of the new theory based on electrical and optical measurements of plasma density 
in femtosecond laser filaments in air and argon. Our results consistently support the established paradigm.   
 
\end{abstract}


\maketitle

Since the first experimental observations of filament propagation in air in the mid-1990s \cite{Brown} 
the field of ultrashort laser filamentation in gases has matured and enabled various technologically important
applications including few-cycle optical pulse generation \cite{few-cycle}, THz generation \cite{THz},
remote sensing \cite{sensing}, laser-triggered lightning \cite{lightning},
and even laser-induced cloud formation \cite{cloud}.
In parallel with experimental progress, the theory and simulation of laser filamentation has advanced 
greatly, allowing for example, for modeling of the supercontinuum generation 
and third-harmonic generation that accompany filament propagation \cite{MiroUPPE}.
Given this tremendous progress, it is still fair to say that the basic paradigm of filament 
formation has remained unchanged from the mid-1990s. Namely, the self-focusing 
collapse of a high peak power input pulse in a gas due to the nonlinear Kerr effect is arrested 
by the defocusing action of free electrons that are generated on the beam axis 
via multi-photon ionization \cite{review1},\cite{review2}.

The established ``old" paradigm has been recently challenged by a suggestion that Kerr self-focusing 
in various gases can saturate and even become defocusing as the optical intensity is increased
\cite{Loriot}. This suggestion represents a radical departure from our past understanding 
of filament formation. Although the saturation of the Kerr effect has been suggested to be a factor in 
laser filamentation \cite{Akozbek,Couairon,Berge}, it was never assumed to be the dominant one. 
If proven valid, the ``new" paradigm would have profound implications in filamentaion science.
On one hand, the sign reversal of the Kerr effect would enable plasma-free filament propagation, 
which could be much longer ranged, owing to the much reduced energy losses into ionization 
compared to that in the established filamentaion scenario \cite{WolfPRL}. 
In addition, various nonlinear optic conversion processes inside filaments, such as third and fifth harmonic generation,
could become very efficient \cite{MiroOL}. On the other
hand, applications that rely on the presence of plasma inside filaments would suffer.

The goal of this paper is to present comprehensive experimental tests of the new paradigm. 
Our results show that defocusing by free electric charges 
is certainly strong enough to balance Kerr self-focusing in femtosecond laser filaments, 
thus supporting the established paradigm. 
Since our experiments are designed such that they measure qualitative, 
not quantitative differences between the predictions of the established and the new theories, 
our conclusions should not be affected by the variability of the material parameters.
We stress that our findings do not imply that the higher-order 
nonlinear terms advocated in \cite{Loriot} are nonexistent. Our conclusion is that,
even if the higher-order nonlinear terms do exist, the free-charge generation and the associated defocusing 
in a filament set in early enough to mask their effect, thus rendering them inoperative.

Our first test is based on the electrical conductivity of filaments.  
Direct and quantitative measurements of the density of charges generated through laser filamentation 
are problematic owing to both the very high intensity inside filaments and short lifetime of the generated plasma.
Accordingly, the reported values of plasma density inside filaments vary by several orders of magnitude 
(see, for example, \cite{review1}). Furthermore, plasma density inside filaments has been shown to be strongly dependent on the 
external focusing conditions \cite{ThebergePRE2006}. Any attempt to validate or disprove the new filamentation paradigm based 
on a quantitative measurement of electric charges generated through filamentation would be very challenging.
Luckily, for a particular pair of gases, namely air and argon, the old and new filamentation theories predict
vastly different values of the generated free-charge density thus offering an opportunity
to validate the new theory based on a semi-qualitative measurement.  

Indeed, by examining the experimental data on the nonlinear refractive index as a function of the 
intensity of light that have been reported for various gases in \cite{Loriot}, one notices that the curves for 
air and argon are essentially identical. This close similarity extends well into the intensity range in which, according to
the new filamentation theory based on this very data, the Kerr effect changes sign and becomes defocusing. 
At the same time, the ionization potential of argon is considerably higher than that of air. 
For laser pulses at 800\,nm that we use in our experiments, it takes 11 photons to ionize argon, 
while ionizing oxygen, one of the major constituents of air, requires only 8 photons.  

According to the above considerations, if the filament is indeed stabilized by sign reversal of the Kerr effect 
(new theory), then filaments generated in air and argon under otherwise identical conditions would have very similar 
spatial intensity distributions. Plasma, which is a bystander in this scenario, would be generated in proportion 
to the multi-photon ionization rate. In that case, one
would expect a lot more plasma to be generated through filamentation in air than in argon. 

\begin{figure}[b]
\includegraphics[width=6.5cm]{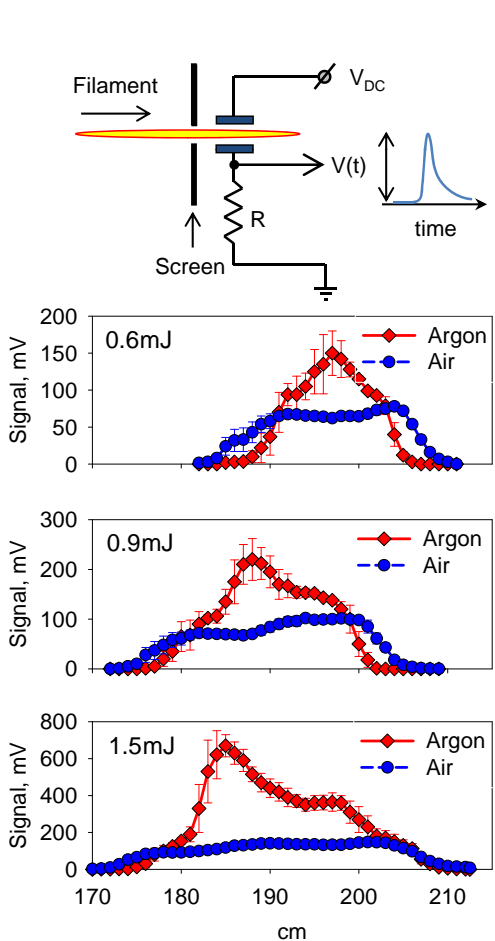}
\caption{\label{fig:1} Top panel: Schematic of the capacitive plasma probe. 
Bottom panels: Experimental data for plasma density generated through filamentation in air and argon
under identical conditions, for three different values of pulse energy.}
\end{figure}

On the other hand, if plasma defocusing is the major stabilizing mechanism in filaments and the Kerr effect is always
positive (old theory), then filamentation in air would generate less plasma than that in argon. 
In this scenario, the higher ionization potential of argon would cause plasma generation 
to be initiated later in the focusing cycle. 
The filament in argon would be thinner than in air. It would take more plasma to defocus this thinner 
filament because it has higher peak intensity and associated stronger self-focusing. 

\begin{figure}[b]
\includegraphics[width=6.5cm]{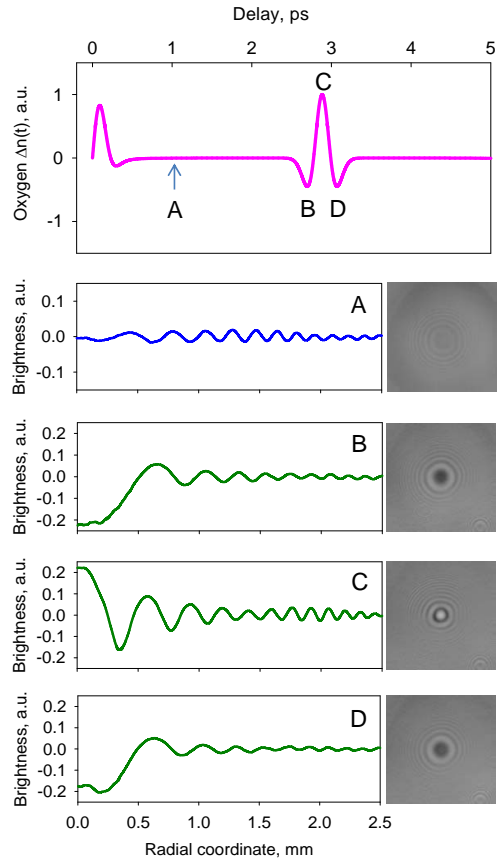}
\caption{\label{fig:2} Top panel: Calculated rotational Raman response for 
molecular oxygen. Bottom panels: Measured diffraction patterns and corresponding radial intensity profiles of the probe pulse 
in air for four particular values of the pump-probe delay as indicated by letters A\,--\,D on the revival curve.
Point A does not overlap with any revival feature of either oxygen or nitrogen.}
\end{figure}

Numerical simulations of filamentaion with 1\,mJ, 30 femtosecond-long pulses in air and argon reported in \cite{WolfPRL}
agree with the above qualitative assessment. For the case of the new theory, these simulations predict about ten times as much
plasma in air as in argon. For the case of the old theory, the simulations predict about twice 
as much plasma in argon as in air.

To experimentally verify the validity of the above predictions, 
we initiate filaments in air and argon under atmospheric pressure, 
using 35 femtosecond-long laser pulses with 800\,nm center wavelength and various pulse energies. 
The laser beam with a 1\,cm diameter is weakly focused with a lens of about 190\,cm focal length, 
making the focusing conditions similar to those used in the simulations reported in \cite{WolfPRL}. 
To measure linear plasma density, in arbitrary units, we use 
a standard capacitive plasma probe schematically shown in the top portion of Figure 1 and described in detail 
elsewhere \cite{Electrodes}. In this particular case, the probe has 1\,cm\,$\times$\,1\,cm square electrodes 
separated by 1.5\,mm. The DC voltage applied to the electrodes is 200\,V.  The results of the plasma density measurements in 
air and in argon are shown in the bottom part of Figure 1. It is evident that in all cases, 
plasma density generated in argon is higher than in air, in agreement with the established filamentation theory. 

\begin{figure}[b]
\includegraphics[width=6.5cm]{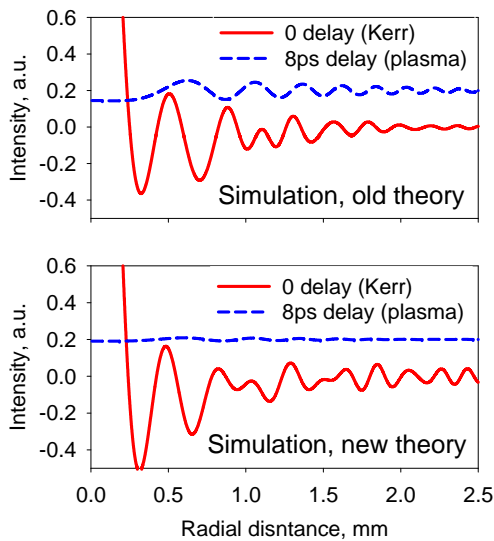}
\caption{\label{fig:4} Radial intensity distributions of the probe pulse in argon that have been numerically 
simulated using old theory (top), and new theory (bottom), for two different values of the pump-probe delay. 
Profiles for 8 picosecond delay are offset upwards, for clarity.} 
\end{figure} 

Owing to the semi-qualitative nature of the above test, the end result confirming the validity of the established theory
appears sufficiently conclusive. However, the weakness of this test is related to the fact that the plasma probe that we use 
is gas specific. Only a small fraction of the generated charges is captured by the probe for each individual laser pulse. 
The measured electrical signal is related not only to the probe geometry and plasma density in the filament, 
but also, in some nontrivial way, to the free-electron mobility and recombination rate in a particular gas. 
Although these parameters are similar
for air and argon, they are certainly not identical for the two gases \cite{gastables}.
Thus strictly speaking, using this plasma probe for quantitative comparison of filament properties 
in different gases is not entirely justified.

\begin{figure}[b]
\includegraphics[width=6.5cm]{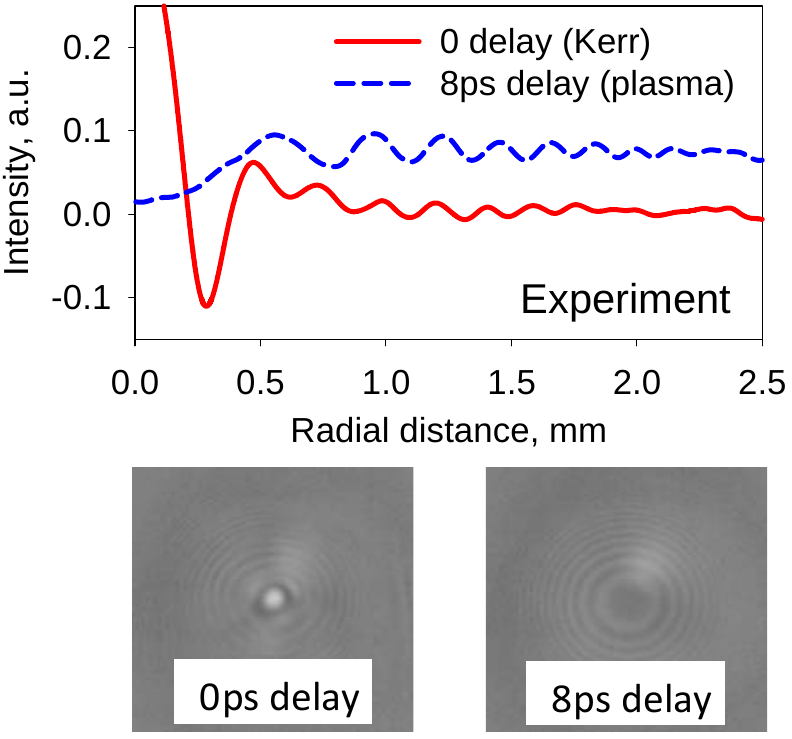}
\caption{\label{fig:5} Experimental data for filamentation in argon under same conditions as those used in simulations 
shown in Figure 3. Profile for 8 picosecond delay is offset upwards. Agreement with the old theory is evident.} 
\end{figure}

Our alternative test of the new theory is based on the diffraction of a collimated probe beam on a plasma channel 
generated through filamentation. The experimental setup is similar to the one reported in \cite{Chindiffraction}. 
Filaments in air and in argon are generated under the same conditions as in the conductivity experiments described above, 
but the pulse energy is fixed at 0.9\,mJ. (At higher pulse energies
in argon we observe the onset of bright conical emission that interfered with the measurements.) 

A fraction of the incident 800\,nm beam is split-off and frequency doubled 
in a 200\,$\mu$m-thick, 1\,cm\,$\times$\,1\,cm BBO crystal and used as a collimated probe beam. 
We estimate that the probe pulse has a duration of less than 50 femtoseconds. The energy of the probe pulse is about 10\,$\mu$J, 
thus the probe propagates in a linear regime. The polarization of the probe is controlled by a half-wave plate, 
and in the experiments reported here polarizations of the pump and probe pulses are parallel. 

The collimated probe beam, after recombination with the pump in a dichroic mirror, propagates collinearly with the pump. 
The time delay between the pump and the probe pulses is controlled via a mechanical delay line.  
The probe beam diffracts on the pump-generated filament and the resulting diffraction pattern is photographed 
by a CCD camera placed at a distance of 75\,cm from the center of the filament. 
The 800\,nm pump light incident on the camera is blocked by a color-glass filter. 
In this setup, the probe pulse non-destructively samples index changes that are experienced 
by the intense pump pulse as it undergoes filamentation.

In air, the nonlinear response of the medium to an ultrashort optical pulse is composed of three components:
The instantaneous Kerr effect, a delayed rotational Raman response which comes in the form of 
periodical revivals, and a free-charge defocusing which lasts for several hundred picoseconds. 
For the purposes of differentiation
between the old and new filamentation theories, our goal here is to establish a relationship between the strengths of 
Kerr self-focusing and free-charge de-focusing. 

The Raman component of the medium response can be calculated \cite{revivals}. The result of this calculation for the 
refractive index change due to the oxygen content of air is shown, in arbitrary units,  
in the top part of Figure 2. 
The bottom part of the Figure shows examples of experimentally recorded diffraction patterns and corresponding
radial intensity distributions of the 400\,nm probe beam, for four particular values of the pump-probe delay. 
The distributions shown are obtained by digitizing the corresponding diffraction
patterns along the line passing through the center of the pattern, followed by subtraction of the profile obtained without 
filament (with the 800\,nm beam blocked).

In all cases, the 
negative (defocusing) swings of the revival curve produce patterns with dark centers, while 
features corresponding to the positive (focusing) swings produce patterns with bright centers.   
Patterns corresponding to the rotational revivals due to molecular nitrogen show the same tendency.   
From this data we deduce an estimate of the temporal resolution  attainable in our experiments. 
Delays between successive focusing-defocusing features on the revival curve (e.g. between points B and C in Figure 2) 
are about 200 femtoseconds. Since our technique clearly distinguishes between these features, 
the temporal resolution of our experiments is substantially better than 200 femtoseconds. The ultimate temporal resolution
that can be achieved is limited by the duration of the probe pulse and by the walk-off between the pump and probe pulses  
due to the group-velocity dispersion over the length of the filament, which is about 20 femtoseconds in this case.  

Pattern A in Figure 2 is obtained for a 1 picosecond delay between pump and probe pulses. 
Neither Kerr effect nor any revival feature for either oxygen or nitrogen 
affects the propagation of the probe pulse in that case, thus this pattern is produced by plasma defocusing. 
As expected, the pattern has a dark center.  
However, quantifying the strength of plasma defocusing and relating it to that of self-focusing based on 
the data obtained for air is problematic because the Kerr response at the zero time delay is masked by the Raman response.

To establish a quantitative relationship between focusing action of the Kerr effect and plasma defocusing, 
we conducted the same diffraction experiment as above in argon atmosphere. Argon is an atomic gas, therefore 
its nonlinear optical response is free from the rotational Raman component inherent to molecular gases such as air. 

Note that changes in refractive index due to the Kerr effect are proportional to the instantaneous optical
intensity. Therefore the transverse geometry of the scatterer that the probe beam diffracts from, 
when pump and probe pulses overlap in time, follows the spatial intensity profile of the pump. 
On the other hand, the spatial profile of the generated plasma is proportional to a high power of intensity. 
Accordingly, the size of the defocusing scatterer due to the plasma is smaller than that of the intensity distribution 
of the pump. As a result, diffraction patterns generated by plasma should be, in general, wider 
compared to those generated by the pump-induced Kerr focusing.

Through extensive numerical simulations of our experiment we found that quantitative comparison between self-focusing 
due to the Kerr effect and the defocusing action of plasma has to be drawn based on the shape of the peripheral parts of the 
diffraction patterns, not their centers. The results of numerical simulations for the intensity distribution of the 
probe beam based on 
the old and new filamentaion theories are shown in Figure 3. In the case of the old model, the oscillations in the peripheral
parts of the patterns corresponding to the zero and large pump-probe delays have comparable contrasts. 
On the contrary, the peripheral oscillation due to plasma defocusing in the case of the new model is
essentially invisible. This again represents a qualitative difference between the predictions of the two theories.

Experimental results for the case of filamentation in argon under the conditions used in the above simulations are
shown in Figure 4. As in the case of filamentation in air, the intensity profiles are obtained by digitizing the
photographed diffraction patterns along the line passing through the pattern's center and subtracting the  
profile obtained with the pump blocked. 
Agreement with the simulation based on the established theory is evident.

In conclusion, we have conducted two independent experiments testing the new filamentation paradigm,
according to which a sign reversal of the Kerr effect is the dominant physical process that stabilizes laser filamentation in gases. 
Both tests consistently disqualified the new paradigm and supported the established theory that treats free-charge defocusing as the
dominant stabilization mechanism.   

The authors acknowledge helpful discussions with Olga Kosareva and See Leang Chin. 
This work was supported by The United States Air Force Office 
of Scientific Research under programs FA9550-10-1-0237 and FA9550-10-1-0561.

$^{*}$ Electronic address: {\it ppolynkin@optics.arizona.edu}

\end{document}